\documentclass[sigconf]{acmart}

\usepackage{booktabs} 
\usepackage{graphicx}
\usepackage{multirow}
\usepackage[ruled,vlined]{algorithm2e}
\usepackage{mathtools}
\usepackage{xspace}
\usepackage{enumerate}
\usepackage[inline,nomargin,marginclue,author=]{fixme}

\setcopyright{acmlicensed}

\acmISBN{} 
\acmDOI{} 

\acmConference[LBRS@RecSys '18]{RecSys 2018}{October 2-7, 2018}{Vancouver, BC, Canada}
\acmYear{2018}
\copyrightyear{2018}

\acmBooktitle{Proceedings of the Late-Breaking Results track part of the Twelfth ACM Conference on Recommender Systems}

\makeatletter
\newcommand{\mathleft}{\@fleqntrue\@mathmargin1pt}
\newcommand{\mathcenter}{\@fleqnfalse}
\makeatother

\newcommand{\algoname}[1]{\textsc{{#1}}\xspace}
\newcommand{\eigensim}{\algoname{EigenSim}}

\newcommand{\itemknn}{\algoname{ItemKNN}}
\newcommand{\slim}{\algoname{Slim}}

\renewcommand{\vec}[1]{\mathbf{#1}}

\begin{document}
\title[Eigenvalue analogy for confidence estimation in recommender systems]{Eigenvalue analogy for confidence estimation in item-based recommender systems}

\author{Maurizio Ferrari Dacrema}
\orcid{0000-0001-7103-2788}
\affiliation{%
  \institution{Politecnico di Milano, Italy}
}
\email{maurizio.ferrari@polimi.it}

\author{Paolo Cremonesi}
\orcid{0000-0002-1253-8081}
\affiliation{%
  \institution{Politecnico di Milano, Italy}
  }
\email{paolo.cremonesi@polimi.it}


\begin{abstract}

Item-item collaborative filtering (CF) models are a well known and studied family of recommender systems, however current literature does not provide any theoretical explanation of the conditions under which item-based recommendations will succeed or fail.

We investigate the existence of an ideal item-based CF method able to make perfect recommendations. This CF model is formalized as an eigenvalue problem, where estimated ratings are equivalent to the true (unknown) ratings multiplied by a user-specific eigenvalue of the similarity matrix.
Preliminary experiments show that the magnitude of the eigenvalue is proportional to the accuracy of recommendations for that user and therefore it can provide reliable measure of confidence.

\end{abstract}

\maketitle

\section{Reliability estimation}

Item-item Collaborative Filtering (CF) is commonly regarded as one of the most effective approaches to build Recommender Systems. 
Given a set of users, a set of items and a set of ratings, item-based CF techniques use available ratings to build relationship between items in the form of an item \textit{similarity matrix}.
Although research on item-based algorithms is not new, the evaluation of their predictive capabilities is purely empirical and still an open issue.
In most works, predictions are expected to be more accurate for items and users associated with many ratings and low variance of ratings. 
Adomavicius et al. \cite{Adomavicius2007Towards} observe that recommendations tend to be more accurate for users and items exhibiting lower rating variance.
Cremonesi et al. \cite{Cremonesi2012User} empirically show that there is some correlation between recall of recommendations and number of ratings in the user profile.
A disadvantage of these approaches is that confidence estimation is based only on user and item ratings, and does not take into account the properties of the prediction model, which could overlook some valuable information.

We show that, under the hypothesis of a \textit{perfect model} (i.e., an oracle able to make perfect recommendations), item-based methods are analogous to an \textit{eigenvalue problem}, where each user profile (i.e., the vector of ratings from a user) is a left eigenvector of the similarity matrix.
Moreover, each user is associated with an eigenvalue, and the magnitude of the eigenvalue is linearly correlated with the accuracy of recommendations for that user.
We call this analogy the \textit{eigenvector analogy}.

\section{Eigenvector Analogy}

A generic item-based model (either CF, CBF or hybrid) predicts the ratings $\vec{\tilde{r}}_u$ of a user $u$ for all items as:
\begin{equation}
  \label{eq:slim}
  \vec{r}_u \vec{S} = \vec{\tilde{r}}_u
\end{equation}
where $\vec{r}_u$ is the profile of the user and $\vec{S}$ is a similarity matrix.
We now assume to have an ideal item-based recommender, able to predict all the user ratings (both known and unknown). 
More formally, this translates into two assumptions:
\begin{description}
\item[Assumption 1: Perfect Predictions.]
Estimated ratings $\vec{\tilde{r}}_u$ are identical to the ratings in the user profile $\vec{r}_u$.
Because of this assumption, we can write $\vec{\tilde{r}}_u = \vec{r}_u$.
\item[Assumption 2: Perfect Knowledge.]
We have the complete knowledge of all the ratings in the user-rating matrix. 
\end{description}
Under these assumptions, the item-based model described by \eqref{eq:slim} 
is analogous to a left eigenvector problem
\begin{equation}
  \label{eq:eigen}
  \vec{r}_u \vec{S} = \lambda_u \vec{r}_u
\end{equation}
where $\vec{r}_u$ is a \textbf{left} eigenvector of matrix $\vec{S}$ and $\lambda_u$ is the corresponding eigenvalue. In the eigenvector formulation \eqref{eq:eigen}, predicted ratings are equivalent to the ratings in the user profile \textit{multiplied by the eigenvalue associated with the user}.
For each user, the corresponding eigenvalue can either flatten out or amplify the differences between predicted ratings.
The closer is an eigenvalue to zero, the more difficult will be for the item-based method to correctly rank items and to distinguish between relevant and non-relevant items.

\subsection{Learning the eigenvalues}
\label{sec:algo}

In order to provide an estimation of the user's eigenvalue, we rewrite \eqref{eq:eigen} in matrix format
\begin{equation}
  \label{eq:Lambda}
  \vec{R} \vec{S} = \vec{\Lambda} \vec{R}
\end{equation}
where $\vec{R}$ is the user-rating matrix, $\vec{S}$ is the similarity matrix and $\vec{\Lambda}$ is a diagonal matrix with eigenvalues on its diagonal.
In order to satisfy the model described by \eqref{eq:Lambda}, we need to find a similarity matrix $\vec{S}$, other than the identity matrix, such that all the user profiles $\vec{r}_u$ in $\vec{R}$ are left eigenvectors of $\vec{S}$.
This problem is called the \textit{inverse eigenvalue problem} 
which, if we also know all the eigenvalues, has the following exact solution: $\vec{S} = \vec{R}^{+} \vec{\Lambda} \vec{R}$.
Where $\vec{R}^+$ is the Moore-Penrose pseudoinverse, which uses singular-value decomposition. We call this model  
\eigensim.
The eigenvalues are then estimated via SGD, optimizing BPR.

\section{Datasets}
\label{sec:results}


We evaluated the performance of \eigensim on two different datasets, namely Movielens 10M (70K users, 10M interactions) and subsample of Netflix (40K users, 1.25M interactions), the dataset used for the Netflix Prize.
The datasets are split by selecting randomly ratings into training (60\%), validation (20\%) and test (20\%) set.

\section{Results discussion}

The main focus of our experiment is to study if there is a correlation between the eigenvalue associated with a user and the quality of recommendations it receives, measured in terms of MAP.
We also investigate if this correlation depends on the specific algorithm used for recommendations. Moreover, as previous work suggested, we study if there is correlation between the user's number of ratings and his/her eigenvalue.
\begin{figure}[t]
  \includegraphics[width=0.35\textwidth]{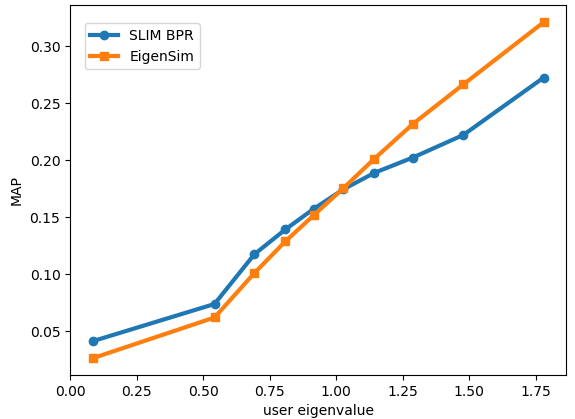}
  \caption{\eigensim and \slim MAP@5 for increasing values of the threshold on $\lambda_u$, on the Movielens 10M dataset.}
  \label{fig:threshold}
\end{figure}
Figure \ref{fig:threshold} presents the performance of \eigensim and \slim on the Movielens 10M dataset. The 70 thousands users of the dataset have been ranked based on their eigenvalue $\lambda_u$ and partitioned into 10 groups of 7K users each. Each point in the figure represents the MAP for each group.
The corresponding eigenvalue is the median of the eigenvalues of the users in the group. The figure clearly shows a linear correlation between the quality of recommendations and the eigenvalue of the users.
When providing recommendations to users with very low eigenvalues the quality of recommendations drops to zero, regardless of the algorithm, while for users with large eigenvalues the quality of recommendations increases for both algorithms.
In terms of MAP, the quality of \itemknn recommendations for users with large eigenvalues ($\lambda_u = 1.75$) is  almost ten times the quality for users with low eigenvalues ($\lambda_u = 0.1$) and twice the average quality across all the users.
It is interesting to observe that eigenvalues affect the quality of recommendations also for an item-based algorithm (\slim) not based on the eigenvalue assumption.

Pearson correlation coefficient between MAP and $\lambda_u$ is 0.99 for both \slim and \eigensim. This confirms that eigenvalues are strong predictors of the quality of recommendations for any item-based algorithms. 
For comparison, on the same dataset, correlation between MAP and profile length is only 0.78 and 0.80 for \slim and \eigensim, respectively. A similar behaviour holds for Netflix.
\begin{figure}[t]
  \includegraphics[width=0.35\textwidth]{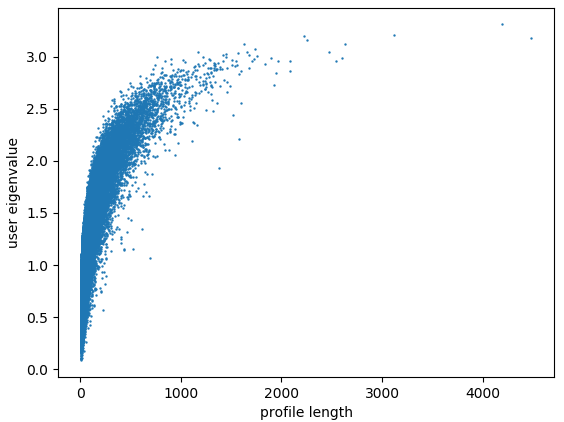}
  \caption{Eigenvalue $\lambda_u$ vs. number of ratings in the user profile for the Movielens 10M dataset.}
  \label{fig:profile}
\end{figure}

Figure \ref{fig:profile} plots, for each user $u$ in the Movielens 10M dataset, the eigenvalue associated with the user, as computed with \eigensim, and the number of ratings in his/her profile. The figures show that there is not a strong correlation between $\lambda_u$ and the number of ratings in the user profile.
For instance, users with a profile length of 50 ratings have eigenvalues ranging from 0.3 (very low quality of recommendations) up to 1.0 (average quality).
The Pearson correlation coefficient between $\lambda_u$ and profile length is 0.76.
The correlation between quality of recommendations and eigenvalues is much stronger than the correlation between quality of recommendations and number of ratings in the user's profile.

These results are not limited to item-based methods but apply also to some model-based matrix factorization methods, such as AsySVD 
and PureSVD
 as, thanks to folding-in, matrix factorization models are equivalent to the FISM item-based method \cite{cremonesi2010performance}.

\section{Conclusion}
\label{sec:conclusions}
We have shown that an ideal item-based method can be formulated as an eigenvalue problem. We show that the magnitude of the eigenvalue is strongly correlated to the accuracy of recommendations for that user and therefore it can provide a reliable measure of confidence. 
Ongoing work is focused on providing a more in-depth theoretical analysis of the eigenvalue analogy and its extension to user-based methods, as well as to validate these results with more datasets and compare against other proposed confidence measures.

\bibliographystyle{ACM-Reference-Format}
\bibliography{bibliography}

\end{document}